# An Empirical Study of Immune System Based On Bipartite Network

Sheng-Rong Zou, Yu-Jing Peng,

Zhong-Wei Guo, Ta Zhou, Chang-gui Gu, Da-Ren He[*]

Yangzhou University, Yangzhou 225002, P. R. China

**ABSTRACT**

**Motivation:** In this paper we present an immune model with bipartite graphs theory. We collect data through COPE database and construct an immune cytokine network. The act degree distribution of this network is proved to be power-law, with index of 1.8. The network is scale-free, with assortativity of –0.27 and similarity of 0.13. The results show that this model could describe the cytokine-cytokine interaction preferably. The basic attributes of this network is well consistent to our preknowledge of immune system.

**Introduction**

Many complex systems use networks as their backbone. Complex Network is a more effective method to describe complex systems. In particular, in biological systems studies, it is increasingly recognized the role played by the topology of cellular networks, the intricate web of interaction among genes, proteins and other molecules regulating cell activity, in unveiling the function and the evolution of living organisms(Jeong et al., 2000; Wagner and Fell, 2001; Jeong et al., 2001; Maslov and Sneppen, 2002; Milo et al., 2002)[1-5]. It is interesting to investigate the network descriptions on immune system (IS). IS is the most important defense system to resist human pathogens. This system consists of immune organs, immune cells and immune molecules. Each cell plays an important role in immune system. The function of many cells in immune system mediated by a group of protein is called cytokines. Cytokines are rapidly produced by immune cells in response to tissue injury, infection, or inflammation. The overproduction of cytokines mediates tissue damage and physiological and molecular mechanisms have evolved to control their production and to prevent the injury during the host response. In addition to regulate cellular interactions, cytokines are the molecular players that signal the brain to respond to the danger of viruses, bacteria, fungi and parasites through an elaborated coordination[6,7]. So we have necessary to find some important immune cells and cytokines .

Although many details of cytokine interactions have been elucidated and the effects of cytokines on a myriad of cellular functions have been described[8], practically nothing is known about the network topological structure of the system as a whole. All cytokine interaction exhibit nonlinear behaviors[9]. In fact, they act in a complex, intermingled network where one cytokine can influence the production of, and response to, many other cytokines. So we believe that IS should be more effectively described by complex network.

In this paper we suggest an immune model with complex network theory [10,11]. The paper is organized as follows. In section two we shall present our model and method. In section three we shall present some results. In the last section, we make a conclusion.

## METHODS

Other authors have modeled the immune system, with a variety of approaches and areas of emphasis [7,12]. But many essential features of this complex system are still not understood. P.Tieri has presented a method to quantifying the relevance of different mediators in the human immune cell network. He draw the cumulative relevance distribution of the immune weighted network , which followed power- law with index of 2.8 [?]. We collect immune data through COPE database and built an immune cytokine network[?].

### 1.1 Bipartite network

Here we use bipartite graphs to built our network. Bipartite network is a graph which connects two distinct sets (or partitions) of nodes, which we will refer to as the top and the bottom set. An edge in the network runs between a pair of a top and a bottom node but never between a pair of top or a pair of bottom nodes[13] (see Fig.1). Typical examples of this type of networks include collaboration networks[14]. Such as the movie-actor, article-author, and board-director network. In the movie-actor network, for instance, the movies and actors are the elements of the top and the bottom set respectively, and an edge between an actor "a" and a movie "m" indicates that "a" has acted in "m". The actors "a "and "a′" are collaborators if both have participated in the same movie ,i.e., if both are connected to the same node "m

'"[13]. The concept of collaboration can be extended to include so diverse phenomena represented by bipartite networks as the city people network, in which an edge between a person and a city indicates that the person has visited that particular city, the word-sentence [15,16], bank-company [17] or donor-acceptor network, which accounts for injection and merging of magnetic field lines[18,19].

As described above is the bipartite network on the knowledge. Next, we used the method of bipartite network to build our immune system. Because the immune system we discuss in this paper have two distinct sets of nodes, the edges between the top nodes and the bottom nodes in our immune system indicates secretion relation.

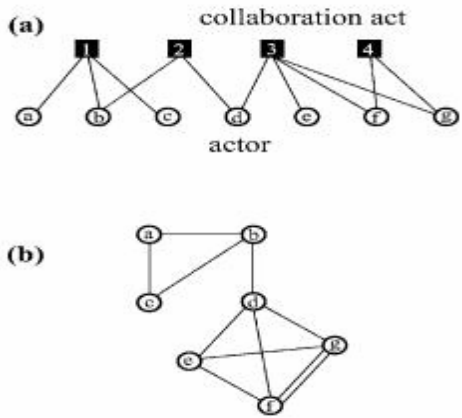

FIG. 1. (a) An example of bipartite graphs. (b) The one-mode projection of the network in （a）.

### 1.2 Our immune network

The immune bipartite network we consider is constituted by two kinds of nodes, one is immune cell types, the other is mediators. In this view, the mediators's role is uniquely one element in a network of mediators interaction. We consider these immune cells mediated by soluble molecules such as cytokines, chemokines, hormones. In the immune system model we construct with 44 top nodes, 1640 bottom nodes, 5391 links. We can also considered top nodes and bottom nodes respectively as act and actor. These data are small compared to other bipartite networks such as the Hollywood actors collaboration network and scientists collaboration network, but it represents a network of very different origin. There have been studies of small-size networks, such as the Jazz instrumentalists collaboration network [17], the underground railway routes network in Boston [18], the airline networks of China and US [19], the power grid network of China and US [4,20], the Indian railway network and the company board member collaboration network [2,22]. One can still obtain valuable statistics about these networks, if the analysis of the data is done properly.

### 2. RESULTS

According to our immune network which we have built, we obtain some results as follows using some algorithm.

## 2.1 Act Degree Distribution

The act degree is an important geometric properties of the complex network. Act degree stand for an actor take part in how many act. In our immune system, act degree can be understood as how many cells secret the same mediator. Fig 2 shows the act degree distribution of the immune system, the distribution can be well described using power–law with index of 1.8. It reveal that there are some nodes with high degree. The biggest act degree is 25. The act degree of two mediators are 25. They are TNF-alpha and IL-8. This shows that 25 kinds of cells can secret the two mediators. The act degree of four kinds mediators is 24. They are TNF-alpha receptors, CCL5, IL-6,IL-2 receptors. The act degree of five kinds mediators is 23. They are TNF-beta receptors, TNF-beta,IL-4 receptors,IL-1 beta, CD54. These mediators are very important in the regulation of immune activity. For example, TNF-alpha, IL-6, IL-8 take part in the inflammatory response which is produced by infection. TGF-beta, IL-6, and so on promote tumor cell growth and metastasis. IL-2 and IFN-γ mediated naturalization reactivity. CCL5 attract immune cells to the local immune response and involved in immune regulation and immunopathological response.

## 2.2 Dyad Act Degree

Dyad act degree is another important properties for bipartite network. Dyad is that two actors and the relationship of them.
Dyad act degree is that a dyad take part in how many act. In our system, we can understood as two mediators can be secreted by how many cells. Fig.3. shows dyad act degree distribution of the immune system, the distribution can be well described by typical SPL(shift power law) [14] functions. The biggest dyad act degree is 22.

### 2.3 Assortativity

In 2002, Newman raised another important parameter of complex network, which is assortativity[20], using a parameter r in between -1 and 1 to describe. Assortativity is that the nodes of the network is whether establish edge links with most similar type other nodes of the network. When r is greater than zero, nodes establish edge links with similar type other nodes of the network, the network is called assortative matching; When r is less than zero , nodes establish edge links with most unsimilar type other nodes of the network, the Network is called disassortative matching. In our Immune System, we draw single assortativity with -0.278609. While multiple edges is counted, we draw multiassortative with 0.143845. The results is consistent to the empirical studies of other system. For example, Newman had the view that social networks are assortative mixing and non-social are disassortative mixing. Some other draw that single assortative of non-social network is lesser than zero and multiassortative is greater than zero.

### 2.4 Similarity

In this paper, we propose a new similarity definition. Similarity of the immune network is the possibilities which each two cells have the common mediators . Its formula is defined as

$$\frac{\sum_i \frac{\sum_i \sqrt{\frac{\sum_j e_{ij} e_{ij}}{k_i k_{i'}}}}{N}}{N}$$

$E_{ij}$ is the adjacency matrix of immune system. "I" and "I'" are cells, and j is mediator. N is the total nodes of the immune system. $K_i$ is the degree of "I" cell. The value of s is between 0 and 1. If s is equal to 1, then the function of all cells are the same. If s is equal to 0, then the function of all cells are not identical. By our computation, the similarity of our immune system is 0.13. It reveals that the possibility of every two different cells secreting the same mediators is only 0.13. This is consisted with our actual immune system. For example, accessory cells only secret Stem Cell Factor; and A549 cells secret Eotaxin and CCL11. This confirmed our guess which any two different cells secreting the same mediators have low possibility. Of course, low possibility does not mean that there is no possibility. Some very important cells can secret the same mediators, and those mediators are also very important in immune system. Such as IL-2, IL-10 ect.

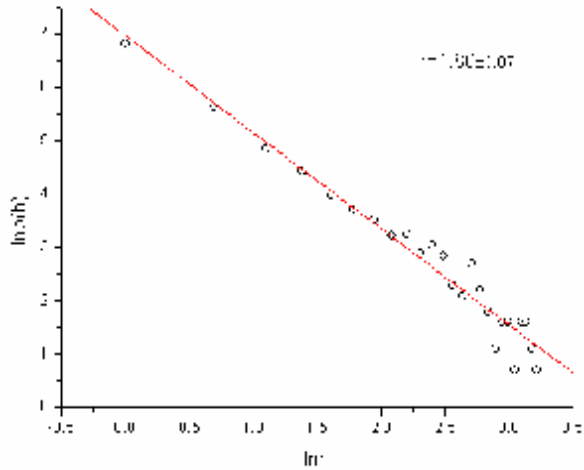

**Fig. 2.**  The graph shows the result of the act degree distribution of immune system. The solid lines represent the least square fitting of the data.

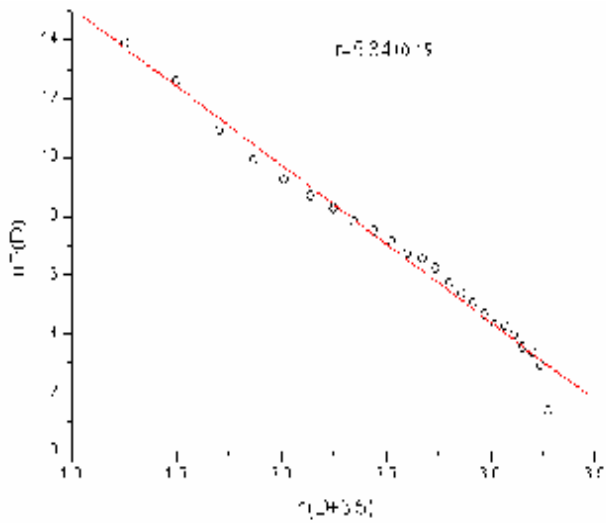

**Fig. 3.** The graph shows the result of the Dyad act degree distribution of immune system. The solid lines represent the least square fitting of the data.

## 3. Conclusion

   From our above analysis, we found that some mediators with high degree are very important mediators in the process of regulating immune activity. For example, like TNF-alpha, IL-8, TNF-alpha receptors, CCL5, IL-6, IL-2 receptors, TNF-beta receptors, TNF-beta, IL-4 receptors, IL-1 beta, CD54 ect. These mediators are important in immune system to regulate their activity. We also found that the assortative of the immune system are negative. It reveals that our immune system is non-social network. Finally we found each two cells similar to a small extent. It reveals that each cell has its unique features. For example, like T cells and B cells, they are all belong to lymphocytes and are main force of non-specific immune system. But the role they played is not the same . T cells implement cellular immunity and B cells implement humoral immunity. The results are well consistent to our preknowledge of immune system．
    The work which we have done is just the beginning. We hope to find more properties of the immune system by complex network analysis.

**ACKNOWLEDGEMENTS**

The research is supported by National Natural Science Foundation of China under the grant No. 70671089 and 10635040.

non-scaling degree distribution in bipartite networks: a numerical and analytical study,